\documentclass[12pt]{article}
\usepackage{amssymb,amsmath,latexsym,natbib,amsthm, easybmat,
  verbatim,graphicx,color,epsfig}
\newcommand{\Var}{\mathsf{Var}}
\newcommand{\Cov}{\mathsf{Cov}}
\newcommand{\tr}{\mathrm{tr}}

\numberwithin{equation}{section}

\newcommand{\proofend}{\hfill$\square$}

\newtheorem{thm}{Theorem}

\begin{document}

\title{Optimal designs for parameters of shifted Ornstein-Uhlenbeck
  sheets measured on monotonic sets}

\author{S\'andor Baran\\
Faculty of Informatics,
University of Debrecen, Hungary,\\
\and
Milan Stehl\'\i k
\\
Department of Applied Statistics,\\
Johannes Kepler University in Linz, Austria}

\date{}
\maketitle

\begin{abstract}
Measurement on sets with a specific geometric shape can be of
interest for many important applications (e.g.
measurement along the isotherms in structural engineering).
In the present paper the properties of optimal designs for estimating
the parameters of shifted Ornstein-Uhlenbeck sheets, that is Gaussian
two-variable random fields with exponential correlation structures, are
investigated when the processes are observed on monotonic sets.
Substantial differences are demonstrated between the cases when
one is interested only in trend parameters
and when the whole parameter set is of interest. The theoretical
results are illustrated by computer experiments and simulated examples
from the field of structure engineering.
From the design point of view the most interesting finding of the paper
is the loss of efficiency of the regular grid design
compared to the optimal monotonic design.
\end{abstract}

\par

\par



\fontsize{10.95}{14pt plus.8pt minus .6pt}\selectfont

\section{Introduction}
   \label{sec:sec1}
Measurement on  sets with a specific geometric shape is of
interest for many important applications, e.g.
measurement along the isotherms.
Starting with the fundamental works of \citet{Hoel58,Hoel61}, the central
importance of equidistant designs for the estimation of
parameters of correlated processes  has been
realized. \citet{Hoel58,Hoel61} compared the efficiencies of equally
spaced designs for one dimensional polynomial models for several
design regions and correlation structures. In this context by a
  design we mean a set $\boldsymbol
  \xi=\{x_1,x_2, \ldots ,x_n\}$ of locations where
  the investigated process is observed. A comparison in a
multi-dimensional setup including correlations can be found in
\citet{Herzberg}. Later \citet{KS} proved that equidistant design
is optimal for estimating the unknown trend parameter  of an
Ornstein-Uhlenbeck (OU) process.  For other papers on
exact designs we refer to the recent
studies on stationary OU process with a constant
mean by \citet{Zagoraiou} and \citet{dkp}. For prediction  of  OU process the
optimality of the equidistant design was proved by
\citet{Zagoraiou2}. Further,
 for a process with a parametrized mean,
it is often possible to find an asymptotic design that performs well
for a large number of design points, see e.g.
\citet{SacksY66,SacksY68}.
All above mentioned papers on optimal design for OU process considered
design region to be an interval from real line. However, a
one-dimensional interval is naturally a directed set induced by total
ordering of real line.
There is a big difference in geometry between plane and line and thus
OU sheet sampled on a two dimensional interval  provides much more
delicate design strategies.

In the present work we derive the optimal exact designs for
parameters of shifted OU sheet measured in the points
constituting a \emph{monotonic set}.
A monotonic set can  be defined in arbitrary Hilbert space $H,$ with
real or complex scalars. For $x, y \in H,$ we
denote by $\langle x, y\rangle $ the real part of the inner product. A set $E
\subset H \times H$ is
called monotonic \citep{Minty1962,Minty1963} provided that for
all $(x_1 , y_1), (x_2 , y_2) \in E$ we
have $\langle x_1 - x_2 , y_1 - y_2 \rangle \geq 0$.
A practical example of such a set are measurements on isotherms of a
stationary temperature field with several applications in thermal slab
modelling (see e.g. \citet{Koizumi} or \citet{babiakMP}). Another
important example in which monotonic measurements appear is motivated by
measuring of methane adsorption \citep{Lee} where keeping all
measurements at isotherm decreases the problems with stability. 
Here we consider the following version of a monotonic
set:

\bigskip
\noindent
{\bf Condition D\/} \ {\em The potential design points
  $\big\{(s_1,t_1),(s_2,t_2),\dots ,(s_n,t_n)\big\}\subset {\mathcal
    X}, \ n\in{\mathbb N}$,  where $\mathcal{X}$ denotes a
    compact design space, satisfy $0<s_1< s_2
< \ldots < s_n$ and $0<t_1< t_2 < \ldots < t_n$.}

We remark that the same observation scheme is used in \citet{bss}
where the authors deal with prediction of OU sheets and derive optimal
designs with respect to integrated mean square prediction error and
entropy criteria.

The paper is organized as follows. In Section \ref{sec:sec2} we introduce the
model to be studied and our notations. Section  \ref{sec:sec3} deals
with an example which motivates this study, namely a design experiment
for measuring on isotherms of a stationary thermal field, while Sections
\ref{sec:sec4}, \ref{sec:sec5}, and
\ref{sec:sec6}  deal with the optimal
designs for the estimation of parameters in our model. We demonstrate the
substantial differences between the cases when only trend parameters
are of interests and when the whole parameter set is of interest. Finally,
Section \ref{sec:sec7} contains some applications and we summarize our
results in Section \ref{sec:sec8}.
To maintain the continuity of
the explanation, the proofs are included in the Appendix.

\section{Statistical Model}
 \label{sec:sec2}

Consider the stationary process
\begin{equation}
   \label{model}
Y(s,t) = \theta +\varepsilon (s,t)
\end{equation}
with  design points taken from a compact design space
$\mathcal{X}=[a_1,b_1]\times [a_2, b_2]$, where $b_1>a_1$ and
$b_2>a_2$ and $\varepsilon (s,t), \ s,t\in {\mathbb R}$, is a
stationary Ornstein-Uhlenbeck
sheet, that is a zero mean Gaussian process with covariance structure
\begin{equation}
   \label{oucov}
{\mathsf E}\,\varepsilon(s_1,t_1)\varepsilon(s_2,t_2)=
\frac{{\widetilde\sigma}^2}{4\alpha\beta}\exp\big
(-\alpha|t_1-t_2|-\beta|s_1-s_2|\big ),
\end{equation}
where $\alpha>0, \ \beta>0, \ \widetilde\sigma>0$. We remark that
$\varepsilon(s,t)$ can also be represented as
\begin{equation*}
\varepsilon (s,t)=\frac{\widetilde\sigma}{2\sqrt{\alpha\beta}}{\mathrm
  e}^{-\alpha t-\beta s}{\mathcal W}\big({\mathrm e}^{2\alpha t},
  {\mathrm e}^{2\beta s}\big),
\end{equation*}
where ${\mathcal W}(s,t), \ s,t\in {\mathbb R}$, is a  standard
  Brownian sheet \citep{Baran, bs}. Covariance structure \eqref{oucov}
  implies that for ${\mathbf d}=(d,\delta), \ d\geq 0, \ \delta \geq 0$,
  the variogram $2\gamma({\mathbf d}):=\Var \big (\varepsilon
  (s+d,t+\delta)-\varepsilon (s,t)\big )$ equals
\begin{equation*}
2\gamma({\mathbf d})=\frac{{\widetilde\sigma}^2}{2\alpha\beta}\Big(1- {\mathrm
  e}^{-\alpha d-\beta \delta}\Big)
\end{equation*}
and the correlation between two measurements depends
on the distance through the semivariogram  $\gamma({\mathbf d})$.

In order to apply the usual approach for design in spatial
modeling \citep{KS} we
introduce $\sigma:=\widetilde\sigma/(2\sqrt{\alpha\beta})$ and instead
of \eqref{oucov} we investigate
\begin{equation}
   \label{oucovmod}
{\mathsf E}\,\varepsilon(s_1,t_1)\varepsilon(s_2,t_2)=
\sigma^2\exp\big (-\alpha|t_1-t_2|-\beta|s_1-s_2|\big ),
\end{equation}
where $\sigma $ is considered as a known parameter. This form
  of the covariance structure is more suitable for statistical
  applications, while \eqref{oucov} fits better to probabilistic
  modelling. Further, we
require {\em  Condition D} to be hold on the design points.
Under {\em Condition D} we may use the construction of \citet{KS} to obtain
the inverse of the covariance matrix of observations which is
tridiagonal. Moreover, in case of an equidistant design the
covariance matrix is Toeplitz.

An exact design allows the experimenter to
plan where to measure the process to optimize a certain measure of
variance of estimators, for optimal design in spatial case see
\citet{Mueller07}. In the literature one can find applications of
various criteria of
design optimality for second-order models. Here we
consider D-optimality, which corresponds to the maximization of objective
function $\Phi(M):=\det (M)$, the determinant of the standard
Fisher information matrix. This method, "plugged" from the widely
developed uncorrelated setup, is offering considerable potential for
automatic implementation, although further development is needed
before it can be applied routinely in practice. Theoretical
justifications for using Fisher information for D-optimal
designing under correlation can be found in \citet{Abt} and
\citet{Pazman07}. \citet{Abt} considered a design space
$\mathcal{X}=[0,1]$ with the covariance structure
$\Cov\big(Y(x),Y(x+d)\big)=\sigma^2e^{-rd}$ and showed that
$\lim_{n\to+\infty}\big(M^{-1}(r,\sigma^2)\big)_{1,1}=0$ and
$\lim_{n\to+\infty}n\big(M^{-1}(r,\sigma^2)\big)_{1,1}=2(r\sigma^2)^2$,
where $\big(M^{-1}(r,\sigma^2)\big)_{1,1}$ denotes the (1,1) entry of
the inverse of the information matrix.

\citet{Zhu} used  simulations (under Gaussian random
field and Mat\'ern covariance structure) to study whether the inverse Fisher
information matrix is a reasonable approximation of the covariance
matrix of maximal likelihood (ML) estimators and a reasonable
design criterion as well. For more references on the Fisher information as
design criterion in the correlated setup see e.g. \citet{Stehlik07}
where the structures of Fisher information matrices for stationary
processes were studied. \citet{Stehlik07} showed that under mild
conditions given on covariance structures
the lower bound for the Fisher information is an increasing function of the
distances between the design points. Particularly, this supports the
idea of increasing domain asymptotics. If for a one-dimensional OU
process only the trend parameters are of interest, then the designs covering
uniformly the whole design space
are very efficient. A similar observation is made in \citet{dkp}
in a more general framework where the authors prove that if
$r\to 0$, then any exact $n$-point D-optimal design in the linear
regression model with exponential semivariogram converges to the
equally spaced design on the one dimensional design interval. A
recurring topic in the recent literature is
that uniform or equispaced designs perform well in terms of
model-robustness when a Bayesian approach is adopted, when the
maximum bias is to be minimized or when the minimum power of the
lack-of-fit test is to be maximized \citep{Goos}. However, an
equidistant design is easy to construct in the case of a single
experimental variable. When more than one variable is involved in an
experiment and the number of observations available is small, it
becomes much more difficult to construct these type of designs.
Uniform design is a kind of space-filling design whose applications
in industrial experiments, reliability testing and computer
experiments is a novel endeavor. The concept of uniform designs was
introduced by \citet{Fang} and has now gained popularity and proved
to be very successful in industrial applications \citep[Chapter
13]{Pham} and in computer experiments \citep{MullerStehlik07,Santner}.

However, for an OU sheet the optimality of a monotonic set is a very
interesting property.
It has become standard practice to select the design points such
as to cover the available space as uniformly as possible, e.g. to
apply the so called space-filling designs. In higher dimensions there
are several ways to produce such designs. The importance of the
discussion whether space-filling designs are superior has been
addressed in the literature recently, see e.g. \citet{PM}. Therein the
review of the circumstances under which this superiority holds is
given together with the clarification of the motives to go beyond
space-filling. In this paper we illustrate that for the OU sheet the
design satisfying monotonicity {\em Condition D} could be  superior
to the space filling grid design.

\begin{figure}[t]
\begin{center}
\leavevmode
\hbox{
\epsfig{file=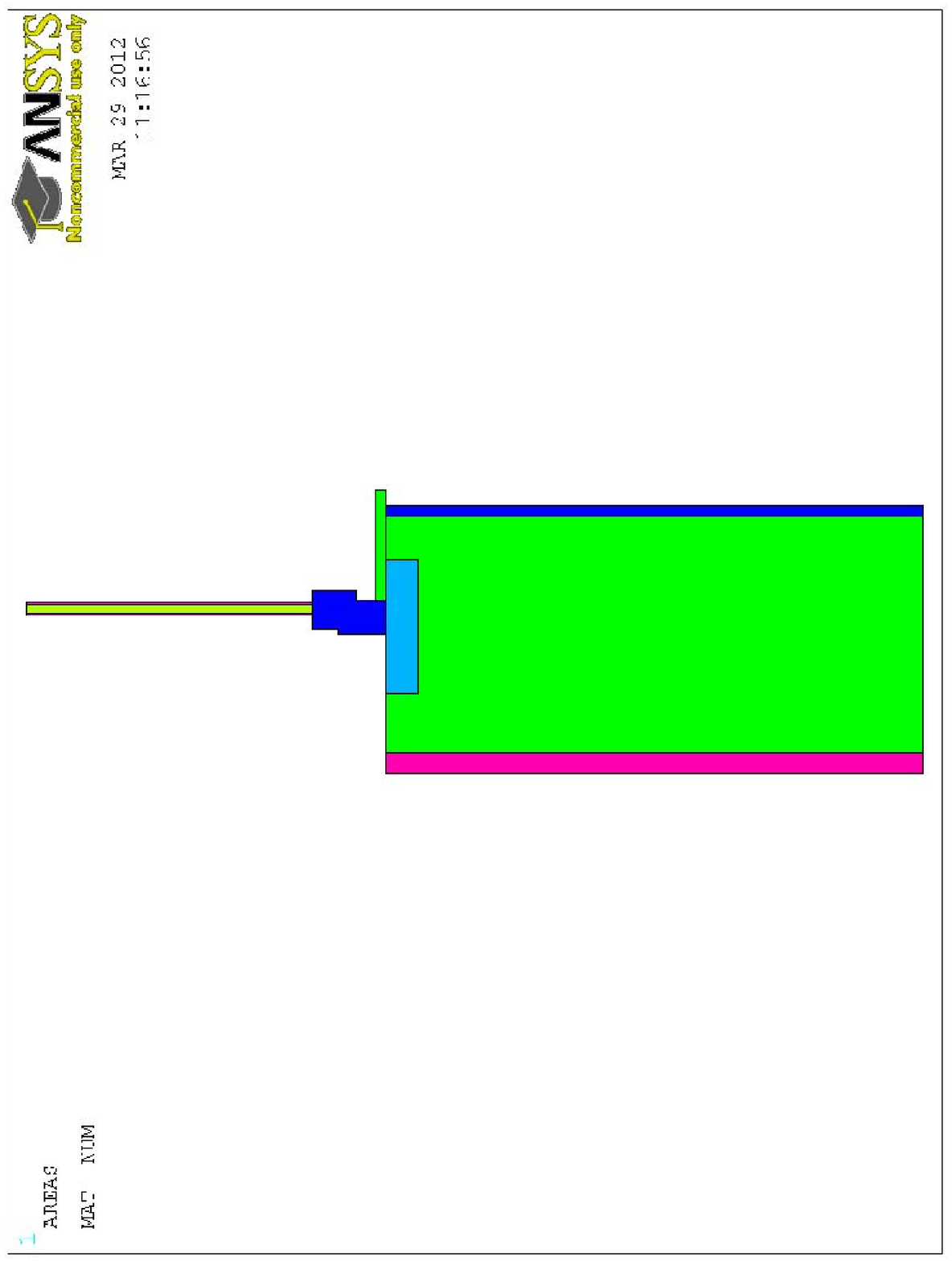,height=6cm, width=8 cm} \quad
\epsfig{file=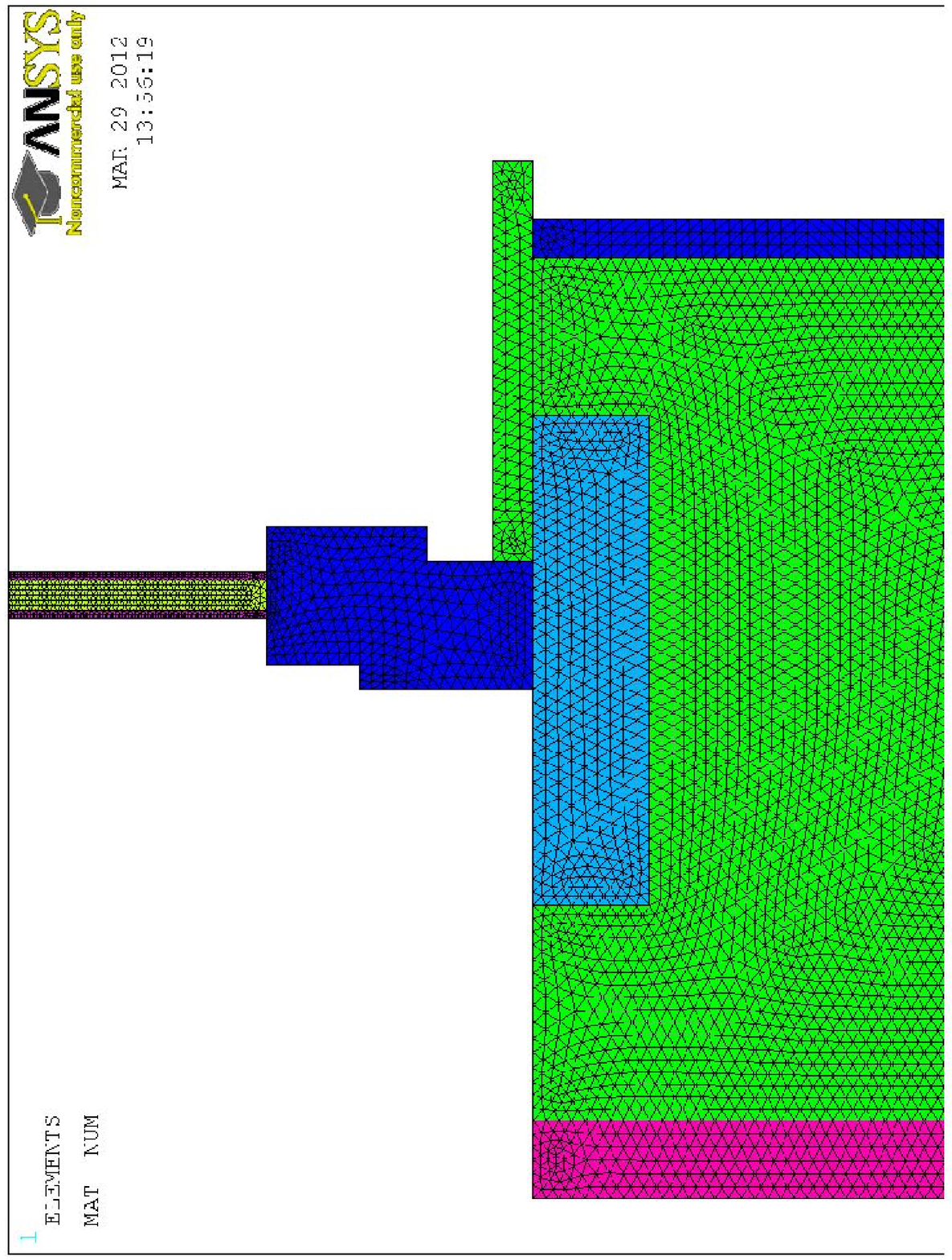,height=6cm, width=8 cm}}

\centerline{\hbox to 9 truecm {\scriptsize (a) \hfill (b)}}

\end{center}
\caption{2D section of a fragment of the building envelope near the
  thermal bridge. (a) Composition of a material; (b) Net for a finite
  element method.}
\label{fig1}
\end{figure}
 The idea of choosing a monotonic set is in particular
  motivated by Markovian properties of the OU sheet.
We are not claiming that monotonic set designs should be used rigidly
in engineering practice, but the aim of our paper is to show that for
OU sheet in some scenarios a monotonic curve could provide better
efficiency than the traditional grid designs.
Therefore, the experimenter is advised to integrate monotonic set design
into his candidate designs portfolio-especially in the cases when there is
a strong intuition/justification of Markovianity of the process.

Being more particular, it is often overseen in practice, that
  information increases with the number of point only in the case of
  independence (or specific form of dependence).
Thus general filling designs, generated without further caution, may increase
variance instead of information \citep{Smit}. Further
discussion on designing for correlated processes in the context of
space filling and its limitations  can be found in
\citet{MullerStehlik07} and \citet{PM}.

Many recent developments on optimal design strategies for
  estimation of parameters should admit that they are mostly a
  \emph{benchmarks}  in the more realistic setups for optimal design
  (like geometric progression ones discussed in in \citet{Zagoraiou}
  for one-dimensional design space, or designs for more complicated
  trends, see e.g. \citet{Chemo}). These
  \emph{benchmarks} should  always be directly confronted with a
  subject science, e.g. with methane modelling in the case of modified
  Arrhenius model as in \citet{Chemo}. In the current paper we provide a
  monotonic design as a benchmark design for a Markovian stochastic
  process measured on rectangle (with continuous time) and a given
  subject science is taken from civil engineering where measurement of
  stationary thermal fields is an issue of interest \citep{Minarova}.

 Finally, we should emphasize that we have not tried to find optimal
 among general design setups -- the main aim of the paper is to
 concentrate entirely on \emph{monotonic set designs}. However,
we are working on finding the optimal regular grid
designs for OU sheets where we are trying to make a comparison (at
least for small sample sizes) with the globally optimal ones.

\section{Motivating example: measurement of a stationary thermal
  field}
  \label{sec:sec3}
Temperature distribution calculation during the process of designing a
building is a necessary part of testing the critical places at the
building envelope. The aim is to increase the minimal surface
temperature, and to predict the possible thermal bridges which are
possible locations of mould growth in the
building. Figure \ref{fig1}a displays the composition of materials of the 2D
 section of a thermal bridge within the building construction, while
 on Figure \ref{fig1}b a net for a finite element method is drawn where
 the points of temperature computation are given. In this way the
 system investigated is a computer experiment
 \citep{SacksSchillerWelch, Santner} modelling the real
 temperature distribution in the building.

\begin{figure}[t]
\begin{center}
\leavevmode
\hbox{
\epsfig{file=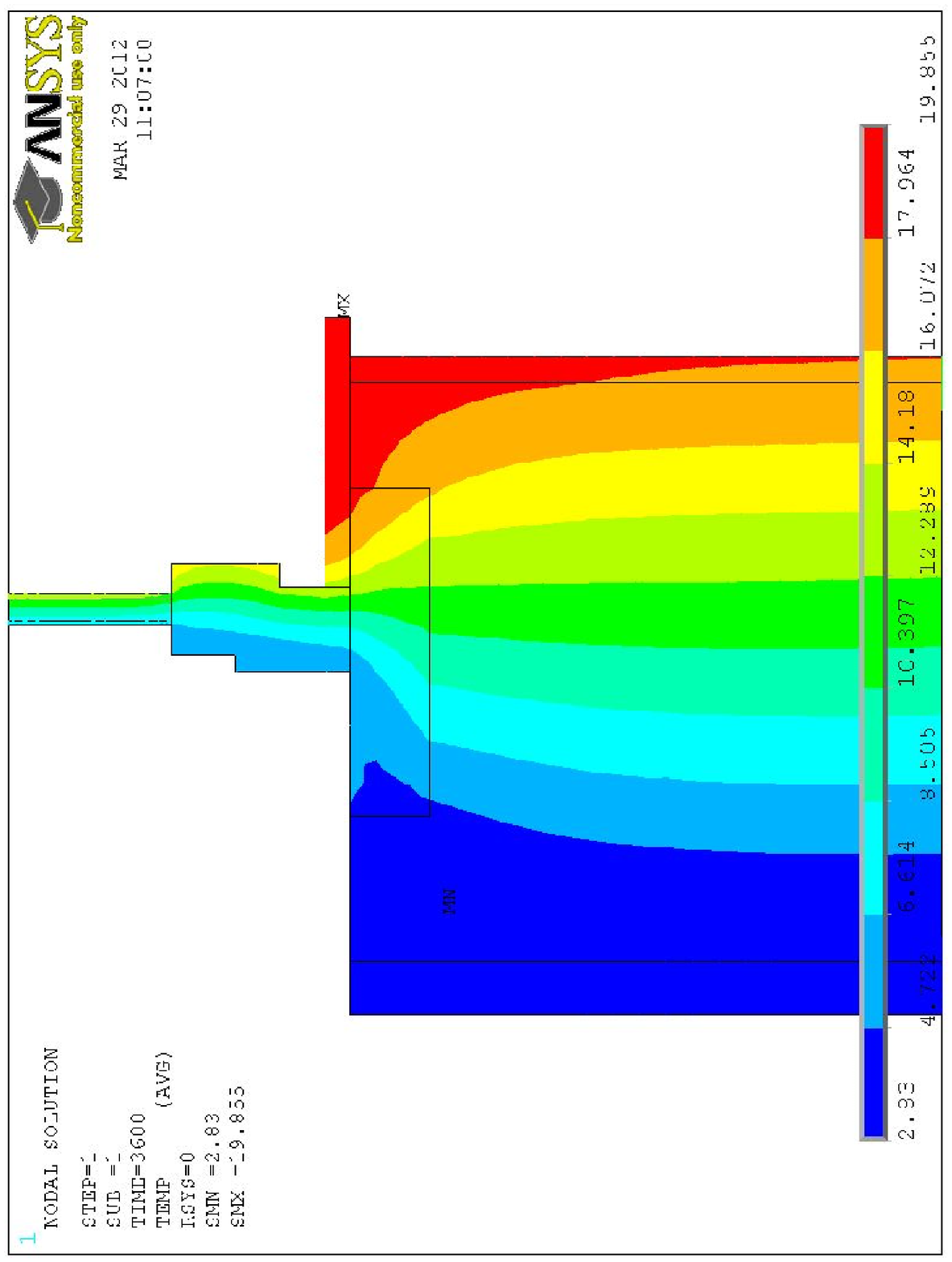,height=6cm, width=8 cm} \quad
\epsfig{file=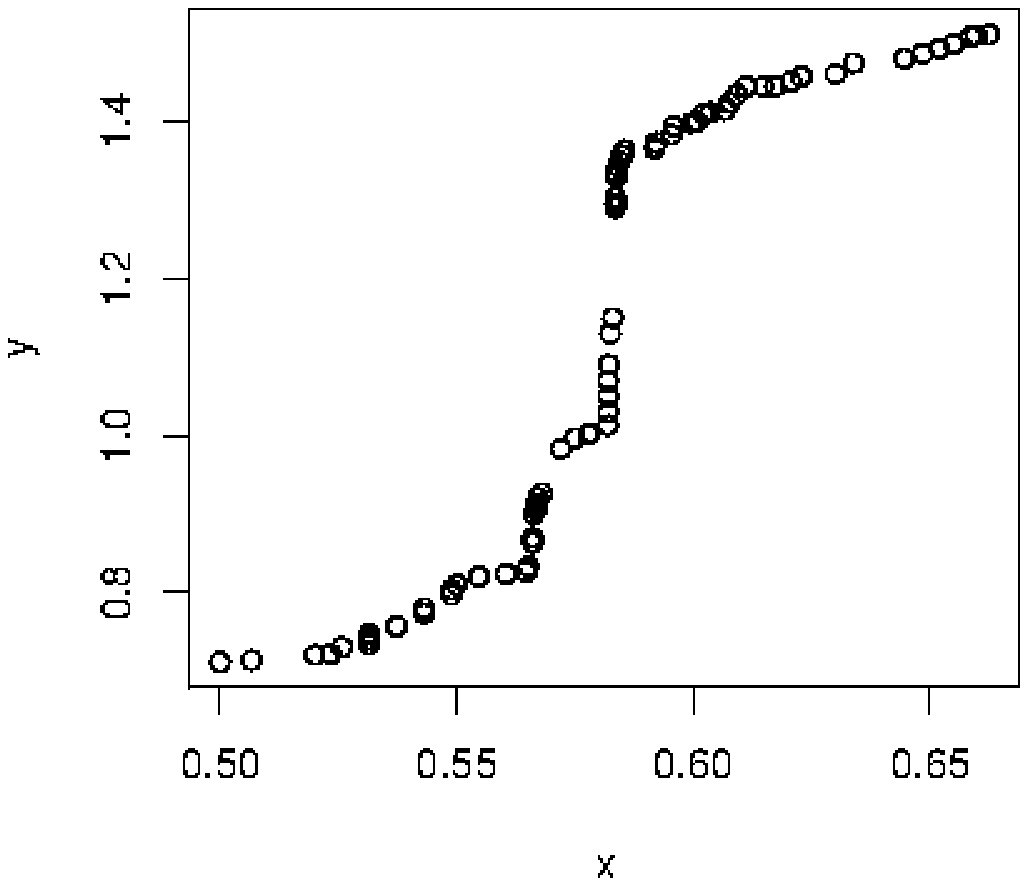,height=6cm, width=8 cm}}

\centerline{\hbox to 9 truecm {\scriptsize (a) \hfill (b)}}

\end{center}
\caption{(a) Isoterms of the thermal field; (b) Observation points on
  Isotherms.}
\label{fig2}
\end{figure}

Data are taken from \citet{Minarova}, where a finite
element method for computation of the temperature field is applied
using software package ANSYS.
Figure \ref{fig2}a illustrates the isotherms of the
thermal field which fit well to measurements in a monotonic set satisfying
{\em Condition D}.

Data points in which we measure the temperature are plotted on
Figure \ref{fig2}b. We assume that the covariance parameters $\alpha$
and $\beta$ are given and we are interested in the estimation of the trend
parameter $\theta $ of model \eqref{model}.
Table \ref{TableMotivatingEx} lists relative
efficiency, information $M_{\theta}$ gained in the data points and the optimal
information gain ($\max M_{\theta}$) of the data from Figure \ref{fig2}b
for three choices of known correlation parameters $\alpha,\beta$. Obviously,
the relative efficiency of the given data points varies with
these parameters.

\begin{table}[b!]
\begin{center}
\begin{tabular}{cccc}
\hline
Correlation Parameters&$M_\theta$&$\max M_\theta$&Efficiency
  ($M_\theta/\max M_\theta$)\\
\hline \hline
$\alpha=\beta=1$&$1.481565$&$1.481695$&$0.99$\\
\hline
$\alpha=1,\beta=10$&$ 4.97261$&$5.081253$&$0.978$\\
\hline
$\alpha=10,\beta=1$&$  2.212449$&$2.212854$&$0.999$\\
\hline
\end{tabular}
\caption{Efficiency depending on correlation parameters.}
\label{TableMotivatingEx}
\end{center}
\end{table}

\section{Estimation of trend parameter only}
   \label{sec:sec4}

Assume first that parameters $\alpha, \beta$ and $\sigma$ of
the covariance structure \eqref{oucovmod} of the OU sheet
$\varepsilon$ are given and we are interested in estimation of the
trend parameter $\theta$. In this case the Fisher information on
$\theta$ based on observations $\big\{Y(s_i,t_i), \ i=1,2,\ldots,
n\big\}$ equals $M_{\theta}(n)={\mathbf 1}^{\top}_nC^{-1}(n,r){\mathbf
  1}_n$, where ${\mathbf 1}_n$ is the column vector of ones of length
$n$,  $r=(\alpha,\beta)^{\top}$, and $C(n,r)$ is the covariance matrix
of the observations  \citep{Pazman07, Xia}.  Further, let $d_i:=s_{i+1}-s_i$ and
$\delta_i:=t_{i+1}-t_i, \ i=1,2, \ldots, n-1$, be the distances between
two adjacent design points.
With the help of this representation one can prove the following theorem.

\begin{thm}
   \label{trend}
Consider the model \eqref{model} with covariance structure
\eqref{oucovmod} observed in points $\big\{ (s_i,t_i),
  \ i=1,2,\ldots ,n\big \}$ satisfying Condition D and assume that the only
  parameter of interest is the trend parameter $\theta$. In this case,
  the equidistant
design of the form $\alpha d_i+\beta {\delta}_i$ to be maximal is optimal
for estimation of  $\theta$.
\end{thm}

According to Theorem \ref{trend} the optimality holds for $\alpha
d_i+\beta \delta_i=\frac{\lambda}{n-1}$, where $\lambda$ is the
``skewed size'' of the design region, i.e. $\lambda:=\alpha
\sum_{i=1}^{n-1} d_i+\beta
\sum_{i=1}^{n-1} \delta_i$ and  $\sum_{i=1}^{n-1}
d_i<b_1-a_1,\ \sum_{i=1}^{n-1} \delta_i <b_2-a_2$. Several situations
may appear in practice.
As now we consider the covariance parameters $\alpha, \beta$ to be
fixed and make inference only on unknown trend parameter $\theta$,
from the proof of Theorem \ref{trend} we obtain
\begin{equation}
  \label{eq:eq3.1}
M_{\theta}(n)=1+\sum_{i=1}^{n-1}\frac{1-q_i}{1+q_i},
\end{equation}
where $q_i:=\exp(-\alpha d_i-\beta \delta _i)$.  Thus, for an optimal
design we have
\begin{equation*}
M_{\theta}(n)=M_{\theta}(n;\lambda)=1+(n-1)\frac{1-\exp(-\lambda
  /(n-1))}{1+\exp(-\lambda/(n-1))},
\end{equation*}
which is an increasing function of both the
number of design points $n$ and the length $\lambda$. Further,
$M_{\theta}(n;\lambda) \to
\lambda/2+1$ as $n\to \infty$ and
$M_{\theta}(n;\lambda) \to n$ as $\lambda\to \infty$, which values are
bounds for information increase in experiments.

To illustrate the latter fact let us consider the design region
${\mathcal X}=[0,1]^2$ and a four-point design, and assume that correlation
parameters are $\alpha=\beta=1$. We are comparing
a regular grid design which puts the four points into the vertices
of the rectangle ${\mathcal X}$ (this design does not satisfy {\em
  Condition D}). The
information corresponding to this design is $M_{\theta}=2.13$.
Having the same design region we cannot reach such an efficiency, because
$\lambda=2$ and $M_{\theta}(n;\lambda) < \lambda/2+1$.
Indeed, the maximal information  gain can be
$M_{\theta}(4;2)=1.965$ which gives us an efficiency of $0.919$.
If we allow the growth of the design region, e.g. ${\mathcal
  X}=[0,x]^2$, for a four-point design,
under the above conditions we obtain
$M_{\theta}=\frac4{1+\exp(-2x)+\exp(-x)}\to 4$ for $x \to \infty$ at a
regular grid design with vertices.

\section{Estimation of covariance parameters only}
   \label{sec:sec5}

Assume now that we are interested only in the estimation of the
parameters $\alpha $ and $\beta$ of the OU
sheet. According to the results of \citet{Pazman07} and \citet{Xia}
the Fisher information matrix on $r=(\alpha,\beta)^{\top}$ has the
form
\begin{equation}
  \label{eq:eq4.1}
M_r(n)=\begin{bmatrix}
        M_{\alpha}(n) &  M_{\alpha,\beta}(n) \\
         M_{\alpha, \beta}(n) &  M_{\beta }(n)
       \end{bmatrix},
\end{equation}
where
\begin{align*}
M_{\alpha}(n)&:=\frac 12 \tr \left\{C^{-1}(n,r)\frac{\partial C(n,r)}{\partial
    \alpha}C^{-1}(n,r)\frac{\partial C(n,r)}{\partial \alpha} \right\}, \\
M_{\beta}(n)&:=\frac 12 \tr \left\{C^{-1}(n,r)\frac{\partial C(n,r)}{\partial
    \beta }C^{-1}(n,r)\frac{\partial C(n,r)}{\partial \beta} \right\}, \\
M_{\alpha,\beta}(n)&:=\frac 12 \tr \left\{C^{-1}(n,r)\frac{\partial
    C(n,r)}{\partial
    \alpha}C^{-1}(n,r)\frac{\partial C(n,r)}{\partial \beta} \right\},
\end{align*}
and  $C(n,r)$ is the covariance matrix of the observations $\big
\{Y(s_i,t_i), \ i=1,2,\ldots, n\big\}$. Note, that here
$ M_{\alpha}(n)$ and $ M_{\beta}(n)$ are Fisher information on
parameters $\alpha$ and $\beta$, respectively, taking the other
parameter as a nuisance.

The following theorem gives the exact form of \ $M_r(n)$ \ for the
model \eqref{model}.

\begin{thm}
    \label{Mrn}
Consider the model \eqref{model} with covariance structure
\eqref{oucovmod} observed in points $\big\{ (s_i,t_i),
  \ i=1,2,\ldots ,n\big \}$ satisfying Condition D. Then
\begin{equation}
   \label{eq:eq4.2}
M_{\alpha}(n)=\sum_{i=1}^{n-1}\frac{d_i^2q_i^2(1+q_i^2)}{(1-q_i^2)^2},
 \quad
M_{\beta}(n)=\sum_{i=1}^{n-1}\frac{\delta_i^2q_i^2(1+q_i^2)}{(1-q_i^2)^2},
\quad
M_{\alpha,\beta}(n)=\sum_{i=1}^{n-1}\frac{d_i\delta_iq_i^2(1+q_i^2)}{(1-q_i^2)^2},
\end{equation}
where $d_i, \delta_i$ and $q_i$ denote the same quantities as in the
previous section, i.e. $d_i:=s_{i+1}-s_i, \ \delta_i:=t_{i+1}-t_i$ and
$q_i:=\exp(-\alpha d_i-\beta \delta_i), \ i=1,2, \ldots ,n-1$.
\end{thm}

Using Theorem \ref{Mrn} one can formulate the following statement on
the optimal design for the
parameters of the covariance structure of the OU sheet.

\begin{thm}
    \label{covpars}
The design which is optimal for estimation of the covariance
parameters $\alpha, \ \beta$ does not exist within the class of
admissible designs.
\end{thm}

\section{Estimation of all parameters}
   \label{sec:sec6}

Consider now the most general case, when both $\alpha,\ \beta$ and
$\theta$ are unknown and the Fisher information matrix on these
parameters equals
\begin{equation*}
M(n)=
\begin{bmatrix}
M_{\theta}(n) & 0 \\
0 &M_r(n)
\end{bmatrix},
\end{equation*}
where $M_{\theta}(n)$ and  $M_r(n)$  are Fisher information matrices on
$\theta$ and $r=(\alpha,\beta)^{\top}$, respectively, see
\eqref{eq:eq3.1} and \eqref{eq:eq4.1}.

\begin{thm}
    \label{allpars}
The design which is optimal for estimation of the covariance parameters
$\alpha$, $\beta$ and of the trend parameter $\theta$ does not exist
within the class of admissible designs.
\end{thm}

Loosely speaking, the optimal designs for the trend have the
  tendency to move the design points as far as possible, while the
  optimal designs for the covariance structure have the tendency to
  shrink the set of design points. However, we can choose a compromise
  between estimating the trend and
correlation parameters. Therefore, similarly to \citet{Zagoraiou}, we may
consider the so-called geometric progression design, which is
generated by the vectors of distances
\begin{equation*}
 {\boldsymbol d}_{n,r_1}:=(k,kr_1,kr_1^2,\ldots,kr_1^{n-2}),\qquad
 {\boldsymbol \delta}_{n,r_2}:=(\ell,\ell r_2,\ell r_2^2,\ldots,\ell r_2^{n-2}),
\end{equation*}
where $0<r_1,r_2\leq 1$.

As $\sum_{i=1}^{n-1}d_i=1$ and $\sum_{i=1}^{n-1}\delta_i=1$, for
$r_1=1,\ r_2=1$ both
constants $k$ and $\ell$ are equal to $(n-1)^{-1}$, while for
$r_1<1$ and  $r_2<1$ we get
$k=\frac{1-r_1}{1-r_1^{n-1}}$ and $\ell =\frac{1-r_2}{1-r_2^{n-1}}$,
respectively. The
tuning parameters $r_1,\ r_2$ can be varied according to the desired
efficiency for the estimation of the trend or the correlation
parameters.

Note, that case $r_1=1,\ r_2=1$ corresponds to the equidistant design,
which we have proved to be optimal for estimation of the trend
parameter, while for $r_1\rightarrow0$, $r_2\rightarrow0$, vectors
${\boldsymbol d}_{n,r_1}$ and ${\boldsymbol\delta}_{n,r_2}$ tend to
the best design for
the estimation of $\alpha$ and $\beta$.

\begin{thm}\label{gpd}
 For any fixed $n>2$, $\alpha>0$, $\beta>0$, the information
 $M_{\theta}(n)$ of the trend is increasing with respect to
 $r_1,r_2$,  while the determinant of the Fisher information $M_r(n)$ of
 covariance parameters has a
 global minimum at $r_1=r_2$.
\end{thm}

\begin{figure}[t]
\begin{center}
\leavevmode
\hbox{
\epsfig{file=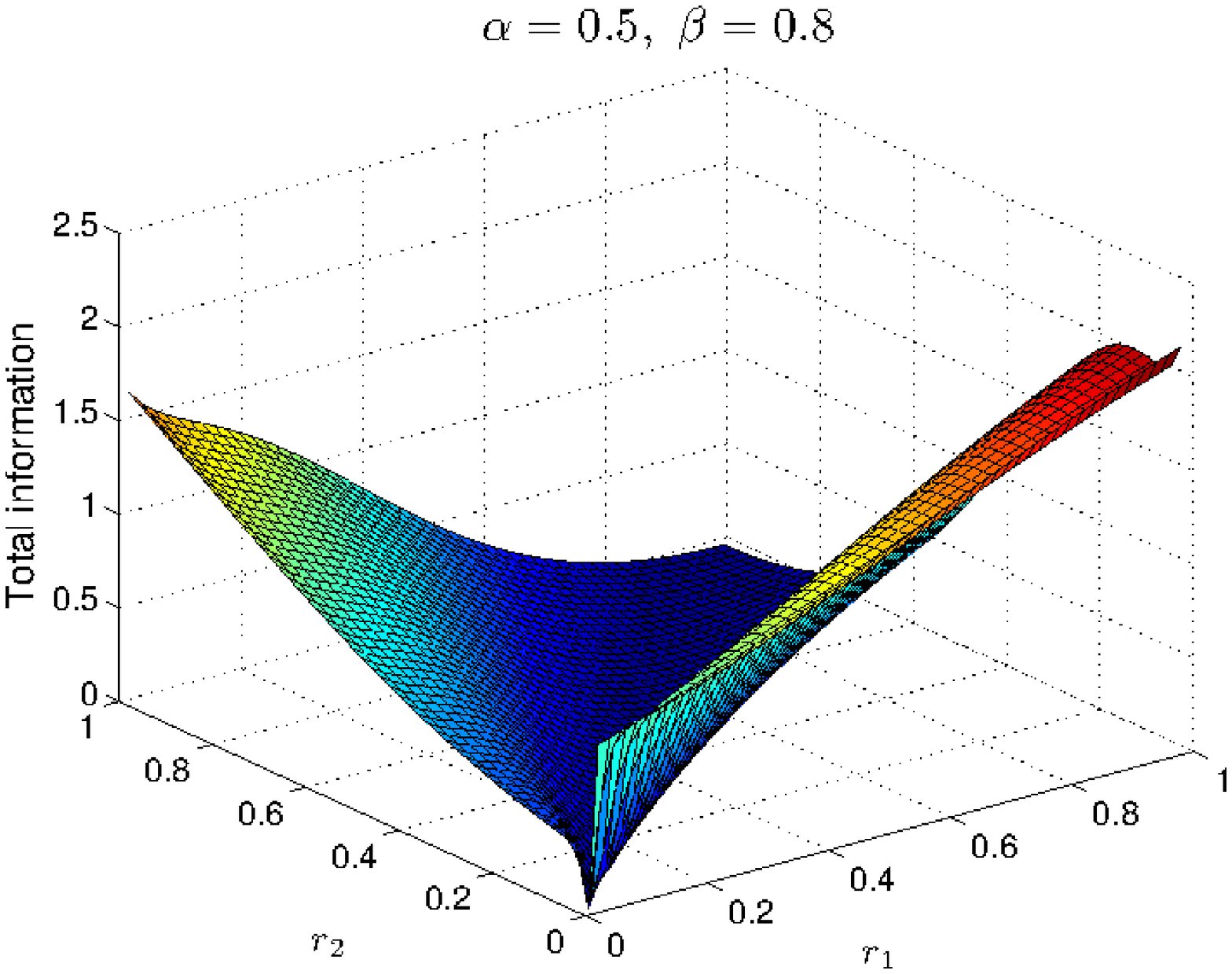,height=6cm} \quad
\epsfig{file=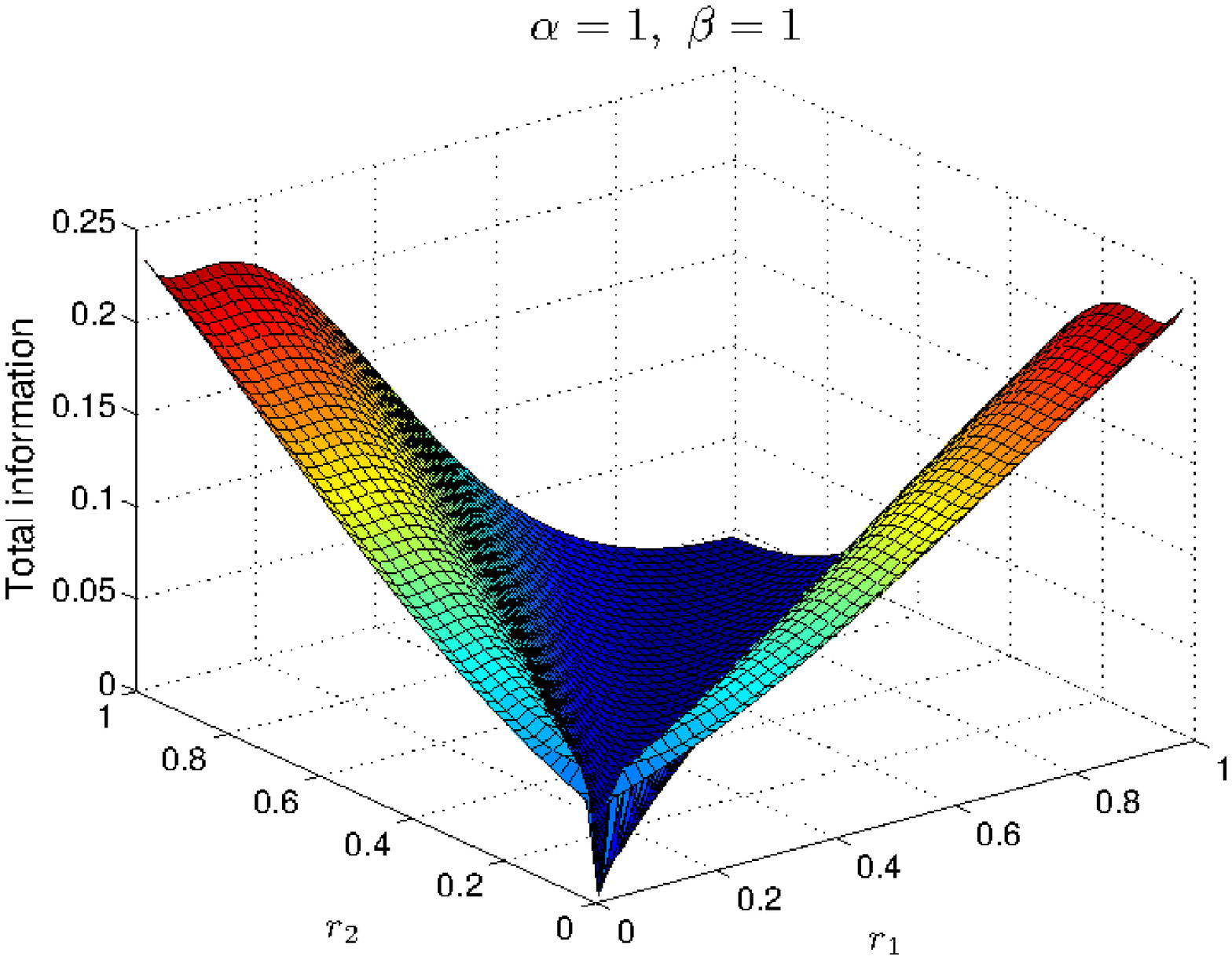,height=6cm}}

\centerline{\hbox to 9 truecm {\scriptsize (a) \hfill (b)}}

\leavevmode
\hbox{
\epsfig{file=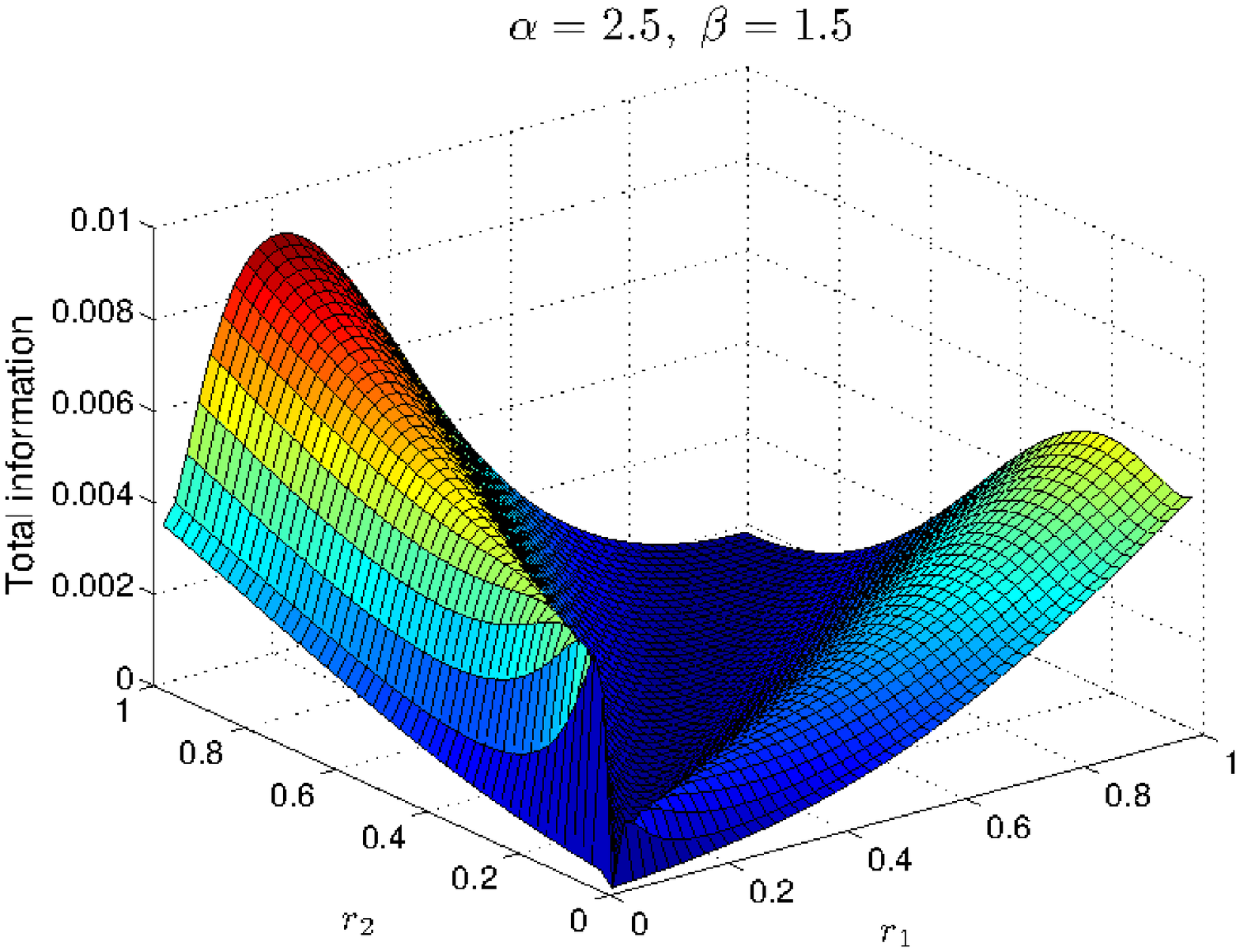,height=6cm} \quad
\epsfig{file=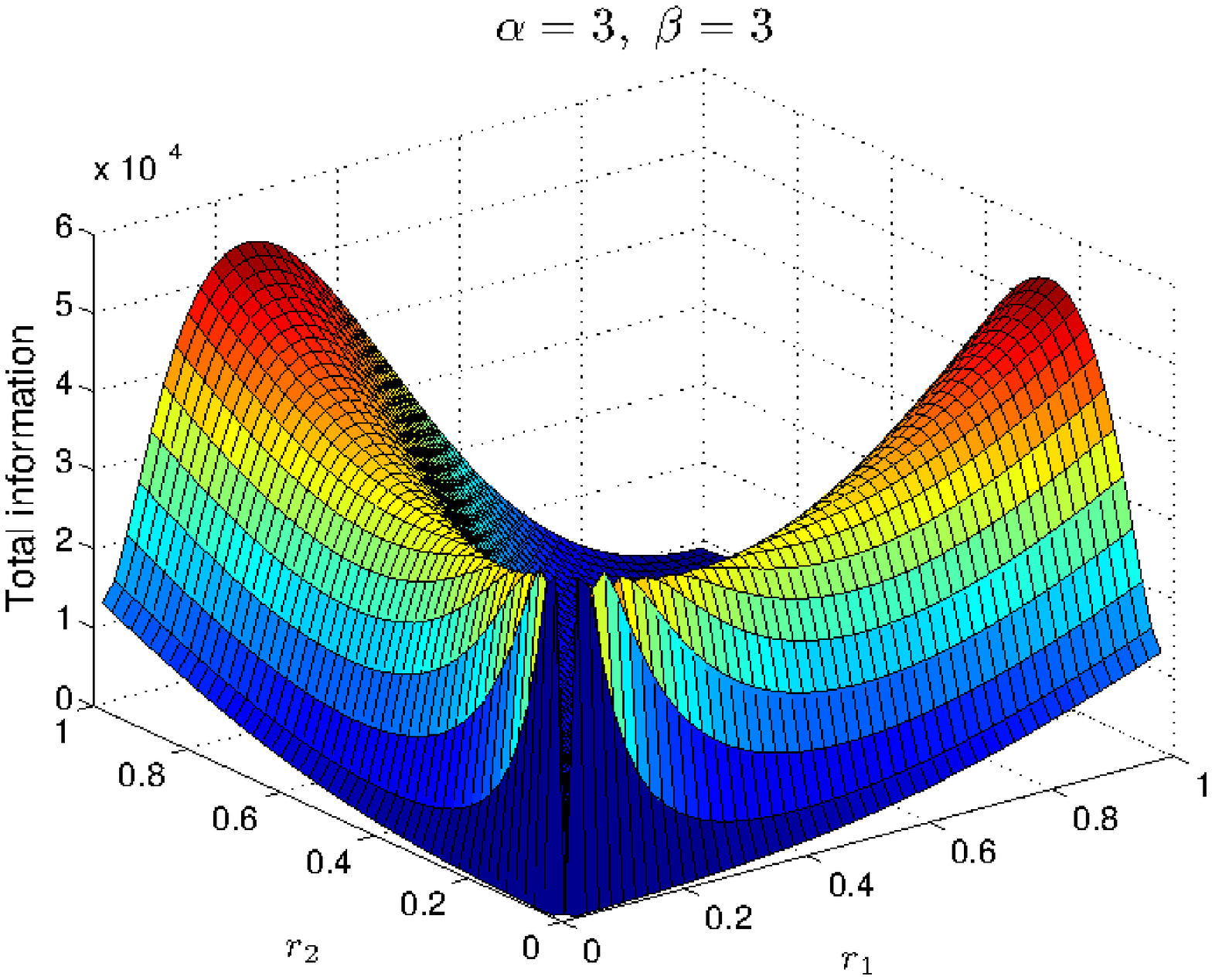,height=6cm}}

\centerline{\hbox to 9 truecm {\scriptsize (c) \hfill (d)}}

\end{center}
\caption{Total information corresponding to an OU process with
  parameters (a) $\alpha=0.5, \ \beta=0.8$; (b) $\alpha=1,\ \beta=1$;
  (c) $\alpha=2.5, \ \beta=1.5$; (d) $\alpha=3,\ \beta=3$.}
\label{totinffig}
\end{figure}

Observe, that Theorem \ref{gpd} obviously implies that the total
information $\det\big(M(n)\big)$
has the same behaviour as $\det \big(M_r(n)\big)$, that is it has a global
 minimum at $r_1=r_2$. This result is clearly illustrated on
 Figure \ref{totinffig}, where for $n=5$ the total information is plotted as a
 function of $r_1$
 and $r_2$ for various combinations of covariance parameters.

\section{Applications to structural engineering}
\label{sec:sec7}

\subsection*{Deterioration of highways}

Typically, engineers are using regular grids for estimation of a random field.
Such an application is demonstrated in \citet{Mohapl} on the data
describing deterioration of a highway in
New York state. Data were collected in four successive years
at distances of $0.2$ miles from each other and form a $4\times 16$
table and based on these data the author estimated the parameters. What is the
efficiency of such a design? The design region has the natural form
$[0,4]\times [0,3.2]$ and the
number of observed points is $64$. In the case  $\alpha=\beta=1$
design satisfying {\em Condition D} and
having 64 points in such a region has
$M_{\theta}(64,7.2)=4.596$.

Now, let us have 16 time coordinates uniformly generated from time region
$[0,4]$ and 16 place coordinates generated from space region
$[0,3.2]$. Then  for time points \ $1.35, \ 3.66, \ 1.86, \ 0.996, \
0.89, \ 1.56, \ 3.37, \ 2.189$,  $0.5157, \ 2.58, \ 0.058, \ 0.32, \
 0.58, \ 1.4, \ 0.36, \ 1.82$ \ and lengths \ $0.64, \ 0.37, \ 1.2, \
 0.91, \ 1.34, \ 2.82, \ 2.56, \ 2.44$,  $0.257, \ 2.568, \ 2.223, \
 0.66, \ 2.298, \ 2.814, \ 2.75, \ 1.61$ \ we obtain $M_\theta=5.2$ in
 the case both parameters $\alpha$ and $\beta$ are equal to $1$.
According to Section \ref{sec:sec4} the maximal information gain with
{\em Condition D}
  for $n=64, \lambda=5.12$ equals
 $3.558592$, thus the relative efficiency is $0.6843446$.
However, there is an open question, how to estimate parameters in such
a set of points, which is far not trivial. Since the
observations form a Gaussian random vector, one can derive the
likelihood function and find the ML
estimates at least numerically. For a regular grid design \citet{ying}
proved consistency and asymptotic normality of the ML estimators, but
according to the authors best knowledge this is the only result in
this direction. The problem is that in
the general case the
dependence of the likelihood function on the parameters and design
points is too complicated to find its asymptotic properties.

When one uses regular grids (for which  \citet{Mohapl} compared least
squares, optimal estimation function and ML estimators) the
following situation occurs: time is measured in 16 equispaced moments
starting from $0$, until $3.75$ by $0.25$, while the deterioration of
the highway is measured in
16 points (by 0.2 miles). Then $M_\theta=4.319177$ (in the case
$\alpha=\beta=1$) with relative efficiency of $0.8275795$.
Table \ref{TableHighway}  is revealing an interesting fact, that regular grid
design (with $256=16^2$ points) has a lost of efficiency with respect  to the
optimal design satisfying {\em Condition D} with the same number of points
in the same design region. This loss can be substantial, dependently
on correlation parameters.

\begin{table}[t!]
\begin{center}
\begin{tabular}{cccc}
\hline
Correlation Parameters&$M_\theta$&$\max M_\theta$&Efficiency
($M_{\theta}/\max M_{\theta}$)\\
\hline \hline
$\alpha=\beta=1$&$4.319177$&$ 4.374803$&$0.987285$\\
\hline
$\alpha=1,\beta=10$&$ 13.13952$&$17.85041$&$0.7360907$\\
\hline
$\alpha=10,\beta=1$&$  14.1108$&$21.20754$&$0.6653671$\\
\hline
\end{tabular}
\caption{Efficiency depending on correlation parameters.}
\label{TableHighway}
\end{center}
\end{table}

\subsection*{Bridge corrosion}

As bridge infrastructures age throughout the world, more
and more bridges are being classified as structurally
deficient \citep{Bhattacharya}. Unfortunately, due to
limited financial resources,
bridge owners are not able to immediately repair or, if
needed, replace all of the structurally deficient bridges in
their inventory. As a result, methods for accurately
assessing a bridge's true load-carrying capacity are needed
so that the limited resources can be spent wisely. Therefore, an
efficient statistical  modeling is needed to
overcome the financial limitations via an efficient design.

\citet{Bhattacharya}
proposed a new model for corrosion rate
that incorporates a multiplicative noise term
ensuring that the corrosion loss function $C(t)$ is non-decreasing
in time, that is  ${\mathrm d}C'(t)\big/{\mathrm
  d}t=\beta(t-T_I)^\gamma\exp(\eta(t))$, for  $t >
T_I$ and $0$ for $t \leq T_I$, where $\beta$ and
$\gamma$ are parameters independent of time and  $\eta(t)$ is
an Ornstein-Uhlenbeck
process. Optimal designs for such a process are derived in \citet{KS}
or \citet{Zagoraiou}. However, several other corrosion sources can be
available yielding a corrosion loss $C$ field depending on two
variables with an error term
$\eta$ forming a planar Ornstein-Uhlenbeck sheet. The  design
strategies studied in this
paper might be of interest for practitioners in estimating the parameters of
such a spatial random field.

\section{Conclusions}
  \label{sec:sec8}

We have constructed exact optimal designs for estimation of parameters of
shifted Ornstein-Uhlenbeck sheets on monotonic sets. The central
importance of equidistant
designs  is visible.
Since the designs strategies for planar OU sheet are much more
difficult than for univariate OU process, the possibility of efficient
designing on a monotonic set is very interesting. We illustrated a
possible  loss of efficiency of the regular grid design with respect
to the
optimal design satisfying monotonicity {\em Condition D}.
A motivation example of isotherm measurement is
given, simulated examples on highway deterioration are also
presented.

In an uncorrelated model the parameter $\sigma$  influences neither the
estimation of the mean value parameters, nor the optimal design. In
the present paper we assume $\sigma$ to be known but a valuable
direction for the future research will be the investigation of models
with unknown nuissance parameter $\sigma$.
\bigskip

\noindent {\large\bf Acknowledgment}

Authors are grateful to Lenka
Filov\'a for her helpful comments during the preparation of the
manuscript. We acknowledge M\'aria Min\'arov\'a for providing us
simulated data of thermal fields.  This research was supported
by the Hungarian Scientific Research Fund under Grants Nos OTKA
T079128/2009 and  OTKA NK101680/2012 and by the Hungarian --Austrian
intergovernmental S\&T cooperation program T\'ET\_{}10-1-2011-0712,
and partially supported by the T\'AMOP-4.2.2.C-11/1/KONV-2012-0001
project. The project was supported by the European Union, with
co-financing from the European Social Fund. The second author acknowledges the
support of the project DESIRE.

\begin{appendix}
\section{Appendix}
  \label{sec:secA}

\subsection{Proof of Theorem \ref{trend}}

According to the notations of Sections \ref{sec:sec3} and
\ref{sec:sec4} let
$d_i:=t_{i+1}-t_i, \ \delta _i:=s_{i+1}-s_i$ and $q_i:=\exp(-\alpha
d_i-\beta \delta _i)$. Similarly to the results of \citet{KS} we have
\begin{equation}
   \label{eq:eqA.1}
C(n,r)=
     \begin{bmatrix}
        1 &q_1 &q_1q_2  &q_1q_2q_3 &\dots &\dots &\prod_{i=1}^{n-1}q_i \\
        q_1 &1 &q_2 &q_2q_3 &\dots &\dots &\prod_{i=2}^{n-1}q_i \\
        q_1q_2 &q_2 &1 &q_3 &\dots &\dots &\prod_{i=3}^{n-1}q_i \\
        q_1q_2q_3 &q_2q_3 &q_3 &1 &\dots &\dots &\vdots \\
        \vdots &\vdots &\vdots &\vdots &\ddots & &\vdots \\
        \vdots &\vdots &\vdots &\vdots & &\ddots &q_{n-1} \\
        \prod_{i=1}^{n-1}q_i &\prod_{i=2}^{n-1}q_i
        &\prod_{i=3}^{n-1}q_i &\dots &\dots &q_{n-1} &1\\
     \end{bmatrix}
\end{equation}
and
\begin{equation}
   \label{eq:eqA.2}
  C^{-1}(n,r)=
     \begin{bmatrix}
        \frac{1}{1-q_1^2} &\frac{q_1}{q_1^2-1} &0 &0 &\dots &\dots &0 \\
        \frac{q_1}{q_1^2-1} &V_{2} &\frac{q_2}{q_2^2-1} &0 &\dots &\dots &0 \\
        0 &\frac{q_2}{q_2^2-1} &V_{3} &\frac{q_3}{q_3^2-1} &\dots &\dots &0 \\
        0 &0 &\frac{q_3}{q_3^2-1} &V_{4} &\dots &\dots &\vdots \\
        \vdots &\vdots &\vdots &\vdots &\ddots & &\vdots \\
        \vdots &\vdots &\vdots &\vdots & &V_{n-1} &\frac{q_{n-1}}{q_{n-1}^2-1} \\
        0 &0 &0 &\dots &\dots &\frac{q_{n-1}}{q_{n-1}^2-1} &\frac{1}{1-q_{n-1}^2}\\
     \end{bmatrix},
\end{equation}
where $V_k:=\frac{1-q_k^2q_{k-1}^2}{(q_k^2-1)(q_{k-1}^2-1)}=\frac
1{1-q_k^2}+ \frac {q_{k-1}^2}{1-q_{k-1}^2}, \ k=2,\dots,n-1$.
Hence, for $M_{\theta}(n)={\mathbf 1}^{\top}_nC^{-1}(n,r){\mathbf
  1}_n$ we obtain
\begin{equation}
   \label{eq:eqA.3}
M_{\theta}(n)=\frac{1-2q_1}{1-q_1^2}+\frac 1{1-q_{n-1}^2}+
  \sum_{i=2}^{n-1}\left(\frac{2q_i}{q_i^2-1}+\frac{1-q_i^2q_{i-1}^2}{(q_i^2-1)
      (q_{i-1}^2-1)}\right)
  =1+\sum_{i=1}^{n-1}\frac{1-q_i}{1+q_i}.
\end{equation}
Now, consider reformulation
\begin{equation*}
M_{\theta}(n)=1+\sum_{i=1}^{n-1}g\big(\alpha
  d_i+\beta \delta_i\big), \qquad \text{where} \qquad
  g(x):=\frac{1-\exp(-x)}{1+\exp(-x)}.
\end{equation*}
As $g(x)$ is a concave function of $x$, by Proposition C1 of
\citet{mo}, $M_{\theta}(n)$ is a Schur-concave function of $\alpha
  d_i+\beta \delta_i, \ i=1,2,\ldots ,n-1$. In this way
  $M_{\theta}(n)$ attains its maximum when $\alpha d_i+\beta
  \delta_i=\lambda /(n-1), \ i=1,2,\ldots ,n-1$, where $\lambda$ is the
  ``skewed size'' of the design rectangle. Hence, an equidistant
 design is the D-optimal for the parameter $\theta$.
 \proofend

\subsection{Proof of Theorem \ref{Mrn}}

By symmetry it suffices to prove
\begin{equation}
   \label{eq:eqA.4}
M_{\alpha}(n)=\frac 12 \tr \left\{C^{-1}(n,r)\frac{\partial C(n,r)}{\partial
    \alpha}C^{-1}(n,r)\frac{\partial C(n,r)}{\partial \alpha}
\right\}=\sum_{i=1}^{n-1}\frac{d_i^2q_i^2(1+q_i^2)}{(1-q_i^2)^2}.
\end{equation}
For $n=2$ equation \eqref{eq:eqA.4} holds trivially. Assume also that
\eqref{eq:eqA.4} is true for some $n$ and we are going to show it for
$n+1$. Let ${\mathbf 0}_{k,\ell}$ be the $k\times \ell$ matrix of
zeros and let
\begin{equation*}
\Delta (n):=\big(-(d_1+d_2+\ldots+d_n)q_1q_2\ldots q_n ,
-(d_2+d_3\ldots+d_n)q_2q_3\ldots q_n, \ldots , -d_nq_n\big)^{\top}.
\end{equation*}
With the help of representation \eqref{eq:eqA.1} one can easily see
that
\begin{equation*}
\frac{\partial C(n+1,r)}{\partial \alpha}=\left[
  \begin{BMAT}{c.c}{c.c}
     \frac{\partial C(n,r)}{\partial \alpha} & \Delta(n)\\
     \Delta^{\top} (n) & 0
  \end{BMAT}
\right],
\end{equation*}
while \eqref{eq:eqA.2} implies
\begin{equation*}
 C^{-1}(n+1,r)=\left[
  \begin{BMAT}{c.c}{c.c}
      C^{-1}(n,r) & {\mathbf 0}_{n,1} \\
     {\mathbf 0}_{1,n} & 0
  \end{BMAT}
\right] +\left[
      \begin{BMAT}{c.c}{c.c}
      \Lambda_{1,1}(n) & \Lambda_{1,2}(n) \\
      \Lambda_{1,2}^{\top}(n) & (1-q_n^2)^{-1}
  \end{BMAT}
     \right],
\end{equation*}
where
\begin{equation*}
\Lambda_{1,1}(n):=\left[
      \begin{BMAT}{c.c}{c.c}
      {\mathbf 0}_{n-1,n-1} & {\mathbf 0}_{n-1,1} \\
      {\mathbf 0}_{1,n-1} & \frac {q_n^2}{1-q_n^2}
      \end{BMAT}
     \right] \qquad \text{and} \qquad
     \Lambda_{1,2}(n):=\left[
      \begin{BMAT}{c}{c.c}
      {\mathbf 0}_{n-1,1} \\
      -\frac {q_n}{1-q_n^2}
      \end{BMAT}
      \right].
\end{equation*}
In this way
\begin{equation*}
C^{-1}(n+1,r)\frac{\partial C(n+1,r)}{\partial \alpha}=\left[
  \begin{BMAT}{c.c}{c.c}
     C^{-1}(n,r)\frac{\partial C(n,r)}{\partial \alpha} &
     C^{-1}(n,r) \Delta(n)\\
     {\mathbf 0}_{1,n} & 0
  \end{BMAT}
\right]+\left[
      \begin{BMAT}{c.c}{c.c}
      {\mathcal K}_{1,1}(n) & {\mathcal K}_{1,2}(n) \\
      {\mathcal K}_{2,1}(n) &  {\mathcal K}_{2,2}(n)
  \end{BMAT}
     \right],
\end{equation*}
with
\begin{alignat*}{2}
{\mathcal K}_{1,1}(n)&:=\left[
      \begin{BMAT}{c}{c.c}
      {\mathbf 0}_{n-1,n} \\
      -\frac {q_n}{1-q_n^2}\big(\Delta^{\top}(n)-(q_n\Delta^{\top}(n-1),0) \big)
      \end{BMAT}
     \right], \qquad
&&{\mathcal K}_{1,2}(n):=\left[
      \begin{BMAT}{c}{c.c}
      {\mathbf 0}_{n-1,1} \\
      -\frac {d_nq_n^3}{1-q_n^2}
      \end{BMAT}
      \right], \\
{\mathcal K}_{2,1}(n)&:=\frac
1{1-q_n^2}\big(\Delta^{\top}(n)-(q_n\Delta^{\top}(n-1),0) \big),
&&{\mathcal K}_{2,2}(n):= \frac {d_nq_n^2}{1-q_n^2}.
\end{alignat*}
Hence,
\begin{align}
   \label{eq:eqA.5}
M_{\alpha}(n+1)=&M_{\alpha}(n)+\tr
\left\{C^{-1}(n,r)\frac{\partial C(n,r)}{\partial \alpha}{\mathcal
    K}_{1,1}(n) \right\}+\tr \left\{C^{-1}(n,r) \Delta(n) {\mathcal
    K}_{2,1}(n) \right\} \\ &+ \frac 12 \tr \left\{{\mathcal K}_{1,1}^2(n)
\right\}+{\mathcal K}_{2,1}(n){\mathcal K}_{1,2}(n)
+ \frac 12 {\mathcal K}_{2,2}^2(n).  \nonumber
\end{align}
After long but straightforward calculations one can get
\begin{align*}
\tr &\left\{C^{-1}(n,r)\frac{\partial C(n,r)}{\partial \alpha}{\mathcal
    K}_{1,1}(n) \right\}=0, \qquad \tr \left\{C^{-1}(n,r) \Delta(n) {\mathcal
    K}_{2,1}(n) \right\}=\frac {d_n^2q_n^2}{1-q_n^2}, \\
\tr &\left\{{\mathcal K}_{1,1}^2(n)
\right\} ={\mathcal K}_{2,1}(n){\mathcal K}_{1,2}(n)={\mathcal
  K}_{2,2}^2(n)=\frac {d_n^2q_n^4}{(1-q_n^2)^2},
\end{align*}
so \eqref{eq:eqA.5} implies
\begin{equation*}
M_{\alpha}(n+1)=M_{\alpha}(n)+\frac{d_n^2q_n^2(1+q_n^2)}{(1-q_n^2)^2},
\end{equation*}
which completes the proof. \proofend
\end{appendix}

\subsection{Proof of Theorem \ref{covpars}}

Consider first the case when we are interested in the estimation of
one of the parameters $\alpha$ and $\beta$ and other parameters are
considered as nuisance.  If $\alpha$ is the parameter of interest
then according to \eqref{eq:eq4.2} the Fisher information on $\alpha$
equals  $M_{\alpha}(n)=\sum_{i=1}^{n-1} F(d_i,\delta_i)$, where
\begin{equation*}
F(d,\delta):= \frac {d^2q^2(1+q^2)}{(1-q^2)^2}\geq 0, \qquad \text{with}
\quad q:=\exp\big(-\alpha d-\beta \delta\big).
\end{equation*}
Due to the separation of the different data points in the expression
of  $M_{\alpha}(n)$ it suffices to consider the properties of the
function $F(d,\delta)$ for $d,\delta\geq 0, \ d\delta \ne0$. Obviously,
\begin{equation}
  \label{eq:eqA.6}
\frac{\partial F(d,\delta)}{\partial d}=\frac
{2dq^2\big((1-q^4)-\alpha d(1+3q^2)\big)}{(1-q^2)^3} \qquad \text{and}
\qquad \frac{\partial F(d,\delta)}{\partial \delta}=\frac
{-2\beta d^2q^2(1+3q^2)}{(1-q^2)^3},
\end{equation}
so the critical points of $F(d,\delta)$ are $(0,\delta), \
\delta>0$. However, at these points the determinant of the Hessian is
zero and for $\delta>0$ we have $F(0,\delta)=0$. Moreover, short
calculation shows that if $d\delta\ne0$ then $F(d,\delta)< 1/(2\alpha^2)$
and $\lim_{\,d,\delta\to 0}F(d,\delta)=1/(2\alpha^2)$.
Hence, the supremum of $F$ is reached at $d=\delta=0$, but
in our context, $d_i\neq0$, $\delta_i\neq0$ for $i=1,2,\ldots, n-1$.

A similar result can be obtained in the case when $\beta$ is the
parameter of interest.

Now, consider the case when both  $\alpha$ and $\beta$ are
unknown. According to \eqref{eq:eq4.1} and  \eqref{eq:eq4.2} the
corresponding objective function to be maximized is
\begin{align}
   \label{eq:eqA.7}
\Phi(d_1,\ldots,d_{n-1},\delta_1,\ldots,\delta_{n-1})=\det\big(M_r(n)\big)
&=\sum_{i=1}^{n-1}\sum_{j=1}^{n-1}(d_i^2\delta_j^2- d_i\delta_id_j\delta_j)
\frac{q_i^2(1+q_i^2)}{(1-q_i^2)^2}\frac{q_j^2(1+q_j^2)}{(1-q_j^2)^2} \\
&=\sum_{i=2}^{n-1}\sum_{j=1}^{i-1}(d_i\delta_j- d_j\delta_i)^2\,
\frac{q_i^2(1+q_i^2)}{(1-q_i^2)^2}\frac{q_j^2(1+q_j^2)}{(1-q_j^2)^2}\geq
0. \nonumber
\end{align}
Obviously, for an equidistant design, where $d_1=\ldots =d_{n-1}$ and
$\delta_1= \ldots =\delta_{n-1}$, the above function equals $0$, that is
this design cannot be optimal. Further,
\begin{align}
  \label{eq:eqA.8}
\frac{\partial \Phi}{\partial
  d_1}&=\frac{2q_1^2(1+q_1^2)}{(1-q_1^2)^2}\big(d_1
\widetilde M_{\beta}(1)-\delta_1\widetilde M_{\alpha ,\beta} (1)\big) -\frac{2\alpha
  q_1^2(1+3q_1^2)}{(1-q_1^2)^3}\big(d_1^2 M_{\beta}(1)+\delta _1^2
\widetilde M_{\alpha}(1) -2d_1\delta_1\widetilde M_{\alpha ,\beta} (1)\big), \\
\frac{\partial \Phi}{\partial
  \delta_1}&=\frac{2q_1^2(1+q_1^2)}{(1-q_1^2)^2}\big(\delta_1
\widetilde M_{\alpha}(1)-d_1\widetilde M_{\alpha ,\beta} (1)\big) -\frac{2\beta
  q_1^2(1+3q_1^2)}{(1-q_1^2)^3}\big(d_1^2 \widetilde M_{\beta}(1)+\delta _1^2
\widetilde M_{\alpha}(1) -2d_1\delta_1\widetilde M_{\alpha ,\beta}
(1)\big), \nonumber
\end{align}
where $\widetilde M_{\alpha}(k), \ \widetilde M_{\beta}(k)$ and
$\widetilde M_{\alpha,\beta}(k), \ k=1,2,\ldots, n-2,$ are the elements
of the Fisher information matrix on  $r=(\alpha,\beta)^{\top}$ corresponding to
observations $\big \{Y(s_i,t_i), \ i=k,k+1,\ldots, n\big\}$ (see
\eqref{eq:eq4.1}), that is
\begin{equation*}
\widetilde M_{\alpha}(k)=\sum_{i=k+1}^{n-1}\frac{d_i^2q_i^2(1+q_i^2)}{(1-q_i^2)^2},
 \quad
\widetilde
M_{\beta}(k)=\sum_{i=k+1}^{n-1}\frac{\delta_i^2q_i^2(1+q_i^2)}{(1-q_i^2)^2},
\quad
\widetilde
M_{\alpha,\beta}(k)=\sum_{i=k+1}^{n-1}\frac{d_i\delta_iq_i^2(1+q_i^2)}{(1-q_i^2)^2},
\end{equation*}
while for $i=2,3,\ldots ,n-1$ we have
\begin{align}
   \label{eq:eqA.9}
\frac{\partial \Phi}{\partial
  d_i}&=\frac{2q_i^2(1+q_i^2)}{(1-q_i^2)^2}\big(d_i
M_{\beta}(i)-\delta_iM_{\alpha ,\beta} (i)\big) -\frac{2\alpha
  q_i^2(1+3q_i^2)}{(1-q_i^2)^3}\big(d_i^2 M_{\beta}(i)+\delta _i^2
M_{\alpha}(i) -2d_i\delta_iM_{\alpha ,\beta} (i)\big), \\
\frac{\partial \Phi}{\partial
  \delta_i}&=\frac{2q_i^2(1+q_i^2)}{(1-q_i^2)^2}\big(\delta_i
M_{\alpha}(i)-d_iM_{\alpha ,\beta} (i)\big) -\frac{2\beta
  q_i^2(1+3q_i^2)}{(1-q_i^2)^3}\big(d_i^2 M_{\beta}(i)+\delta _i^2
M_{\alpha}(i) -2d_i\delta_iM_{\alpha ,\beta} (i)\big). \nonumber
\end{align}
Solving recursively the equations \eqref{eq:eqA.9} under the
assumption $d_i\delta_i\ne 0, \ i=1,2,\ldots ,n-1,$ for the critical
points of $\Phi$ we obtain relations
\begin{equation}
  \label{eq:eqA.10}
\frac {d_i}{d_1}=\frac {\delta _i}{\delta_1}=:c_i>0, \quad \text{that
  is} \quad d_i=c_id_1, \ \delta_i=c_i\delta _1, \qquad i=1,2,\ldots ,n-1.
\end{equation}
These solutions also solve \eqref{eq:eqA.8} and short calculations
show that for all $d_1,\delta_1,c_1,\ldots ,c_{n-1}>0$ we have
$\Phi (d_1,c_1d_1, \ldots ,c_{n-1}d_1,\delta_1,c_1\delta_1,\ldots
,c_{n-1}\delta_1)=0$. Hence, critical points determined by
\eqref{eq:eqA.10} are minimum points of $\Phi$.
Thus, the maximum of $\Phi(d_1,\ldots,d_{n-1},\delta_1,\ldots,\delta_{n-1})$
can only be attained at the boundary points, but in our context,
$d_i\notin\{0, b_1-a_1\}$ and $\delta_i\notin\{0, b_2-a_2\}$. \proofend

\subsection{Proof of Theorem \ref{allpars}}

As $\det\big(M(n)\big)=M_{\theta}(n)\det\big(M_r(n)\big)=M_{\theta}(n)\Phi$,
according to \eqref{eq:eqA.3} and  \eqref{eq:eqA.7}, for unknown
parameters $\alpha, \ \beta$ and $\theta$ the
objective function to be maximized is
\begin{align}
   \label{eq:eqA.11}
\Psi(d_1,\ldots,d_{n-1},\delta_1,\ldots,\delta_{n-1})=
\left( \frac 2{1+q_1}\!+\!\sum_{i=2}^{n-1}\frac{1-q_i}{1+q_i}\right)\left
  (\sum_{i=2}^{n-1}\sum_{j=1}^{i-1}(d_i\delta_j\!-\! d_j\delta_i)^2\,
\frac{q_i^2(1+q_i^2)}{(1-q_i^2)^2}\frac{q_j^2(1+q_j^2)}{(1-q_j^2)^2}
\right)
\end{align}
For $d_1=\ldots =d_{n-1}$ and $\delta_1= \ldots =\delta_{n-1}$,
we have $\Phi(d_1,\ldots,d_{n-1},\delta_1,\ldots,\delta_{n-1})=0$,
thus an equispaced design cannot be optimal.

Further,
\begin{equation*}
 \frac{\partial \Psi}{\partial d_i}=
 M_{\theta}(n)\frac{\partial\Phi}{\partial d_i}
-\frac{2\alpha q_i}{(1+q_i)^2}\Phi \qquad \text{and} \qquad
 \frac{\partial \Psi}{\partial
   \delta_i}=M_{\theta}(n)\frac{\partial\Phi}{\partial
   \delta_i}-\frac{2\beta q_i}{(1+q_i)^2}\Phi,
\end{equation*}
where the expressions for  $\partial\Phi/\partial d_i$ and
$\partial\Phi/\partial\delta_i$ are given by \eqref{eq:eqA.10}.
Solving the above equations for the critical points of $\Psi$ we
obtain the relations \eqref{eq:eqA.10}.
However, $\Psi (d_1,c_1d_1, \ldots
,c_{n-1}d_1,\delta_1,c_1\delta_1,\ldots,c_{n-1}\delta_1)=0$,
thus the function $\Psi$ attains its minimum at the points determined
by \eqref{eq:eqA.10}. \proofend

\subsection{Proof of Theorem \ref{gpd}}
Consider first $M_{\theta}(n)$ and according to \eqref{eq:eq3.1}
\begin{equation*}
M_{\theta}(n)=M_{\theta}(n;r_1,r_2)=1+\sum_{i=1}^{n-1}
f\big(d_i(r_1),\delta_i(r_2)\big), \quad \text{where} \quad
f(d,\delta)=\frac{{\mathrm e}^{\alpha
  d+\beta \delta}-1}{{\mathrm e}^{\alpha d+\beta \delta}+1}.
\end{equation*}
Obviously, for $r_1=1,\ r_2=1$, the geometric
progression design corresponds to the equidistant design, which is
optimal for the estimation of the trend parameter.
Let $0<r_1, r_2 <1$ and one has to prove that
\begin{equation*}
 \frac{\partial M_{\theta}(n;r_1,r_2)}{\partial r_1}=\sum_{i=1}^{n-1}\frac{\partial
   f(d_i(r_1),\delta_i(r_2))}{\partial d}\frac{\partial
   d_i(r_1)}{\partial r_1}>0 \ \ \text{and} \ \
 \frac{\partial M_{\theta}(n;r_1,r_2)}{\partial r_2}=\sum_{i=1}^{n-1}\frac{\partial
   f(d_i(r_1),\delta_i(r_2))}{\partial \delta}\frac{\partial
   \delta_i(r_2)}{\partial r_2}>0.
\end{equation*}
Now,
\begin{equation*}
\frac{\partial f(d,\delta)}{\partial d}=\frac{2\alpha{\mathrm e}^{\alpha
  d+\beta \delta}}{({\mathrm e}^{\alpha d+\beta \delta}+1)^2}>0,
\end{equation*}
which, as a function of $\alpha d+\beta \delta$,  is strictly
decreasing. In this way we can use the arguments
of Proof of Theorem 5.1 of \citet{Zagoraiou},
where a one-dimensional OU process is investigated. From
$d_i(r)=\delta_i(r)=\frac{(1-r)r^{i-1}}{1-r^{n-1}},\ i>1,$ we obtain
$d_1(r_1)>\ldots >d_{n-1}(r_1)$ and $\delta_1(r_2)>\ldots
>\delta_{n-1}(r_2)$ which implies
\begin{equation*}
0<\frac{\partial f\big(d_1(r_1),\delta_1(r_2)\big)}{\partial d}<
\ldots < \frac{\partial
  f\big(d_{n-1}(r_1),\delta_{n-1}(r_2)\big)}{\partial d}.
\end{equation*}
Further,
\begin{equation*}
\frac{\partial d_{i}(r_1)}{\partial
  r_1}=\frac{r_1^i}{(r_1-r_1^n)^2}\big((r_1^{n-1}(n-i)-r_1^n(n-i-1)+i-1-r_1i\big),
\qquad i =1,2, \ldots,n-1,
\end{equation*}
and due to $\sum_{i=1}^{n-1}d_i(r_1)=1, \ 0<r_1\leq1$, we have
$\sum_{i=1}^{n-1}\frac{\partial d_i(r_1)}{\partial r_1}=0$.
Now,  let $j$ be
the smallest integer such that $\frac{\partial d_{i}(r_1)}{\partial
  r_1}\geq 0$ for
$i=j,\ldots,n-1$, and according to \citet{Zagoraiou} such integer
exists. Then
\begin{align*}
\sum_{i=1}^{n-1}\frac{\partial
   f(d_i(r_1),\delta_i(r_2))}{\partial d}\frac{\partial
   d_i(r_1)}{\partial r_1}=&\sum_{i=1}^{j-1}\frac{\partial
   f(d_i(r_1),\delta_i(r_2))}{\partial d}\frac{\partial
   d_i(r_1)}{\partial r_1}+\sum_{j=1}^{n-1}\frac{\partial
   f(d_i(r_1),\delta_i(r_2))}{\partial d}\frac{\partial
   d_i(r_1)}{\partial r_1} \\ >&\frac{\partial
   f(d_j(r_1),\delta_j(r_2))}{\partial d}\sum_{i=1}^{n-1}\frac{\partial
   d_i(r_1)}{\partial r_1}=0.
\end{align*}
The positivity of the other partial derivative of
$M_{\theta}(n;r_1,r_2)$ can be proved exactly in the same way.

Finally, the second statement of the theorem is a direct consequence
of \eqref{eq:eqA.7}, since if $r_1=r_2$ then for all $i=2,3, \ldots
,n-1$ and $j=1,2, \ldots ,i-1$ we have
$d_i(r_1)\delta_j(r_2)-d_j(r_1)\delta_i(r_2)=0$.
\proofend

\end{document}